\documentclass[12pt]{iopart}

  \expandafter\let\csname equation*\endcsname\relax
  \expandafter\let\csname endequation*\endcsname\relax
\usepackage[T1]{fontenc}
\usepackage{color}
\usepackage{amsmath}
\usepackage{amssymb}
\usepackage{graphicx}
\usepackage{iopams}
\usepackage{esint}
\usepackage{ulem}   % to strike things out
\makeatletter
\@ifundefined{textcolor}{}
{%
 \definecolor{BLACK}{gray}{0}
 \definecolor{WHITE}{gray}{1}
 \definecolor{RED}{rgb}{1,0,0}
 \definecolor{GREEN}{rgb}{0,1,0}
 \definecolor{BLUE}{rgb}{0,0,1}
 \definecolor{CYAN}{cmyk}{1,0,0,0}
 \definecolor{MAGENTA}{cmyk}{0,1,0,0}
 \definecolor{YELLOW}{cmyk}{0,0,1,0}
}

\@ifundefined{definecolor}
 {\usepackage{color}}{}
\@ifundefined{definecolor}{\usepackage{color}}{}
\@ifundefined{definecolor}{\usepackage{color}}{}
\@ifundefined{definecolor}{\usepackage{color}}{}

\newcommand{\vect}[1]{{\boldsymbol{#1}}}

\makeatother

\begin{document}

\title{Unparticle mediated superconductivity}

\author{James P. F. LeBlanc$^{1,2}$ and Adolfo G. Grushin$^1$}
%\affiliation{Max-Planck-Institut f\"{u}r Physik komplexer Systeme, 01187 Dresden, Germany}
%\affiliation{Department of Physics, University of Michigan, Ann Arbor, Michigan 48109, USA}
\address{$^1$Max-Planck-Institut f\"{u}r Physik komplexer Systeme, 01187 Dresden, Germany}
\address{$^2$Department of Physics, University of Michigan, Ann Arbor, Michigan 48109, USA}

%\author{Adolfo G. Grushin}
%\address{Max-Planck-Institut f\"{u}r Physik komplexer Systeme, 01187 Dresden, Germany}

\ead{jpfleblanc@gmail.com}
\date{\today}

\begin{abstract}
In this work we introduce the possibility of unparticle mediated superconductivity.
We discuss a theoretical scenario where it can emerge and
show that a superconducting state is allowed by deriving and solving the gap equation for $s$-wave pairing of electrons 
interacting through the unparticle generalization 
of the Coulomb interaction.  
The dependence of the gap equation on the unparticle energy scale 
$\Lambda_{U}$ and the unparticle scaling dimension $d_{U}$ enables us to find a richer set of solutions
compared to those of the conventional BCS paradigm. 
We discuss unconventional features within this construction, including the resulting insensitivity of pairing to the density of states 
at the Fermi energy for $d_{U}=3/2$ of the superconducting gap and suggest possible experimental scenarios for this mechanism. 
\end{abstract}

\maketitle

\section{Introduction} 
One of the most remarkable consequences of the discovery of unconventional superconductivity (SC)
in both cuprates and pnictides~\cite{bednorz:1986, kamihara:2008} is, arguably, the richness of theoretical ideas that have been put forward to understand this
physical phenomenon \cite{N11,HKM11}. Searching for novel mechanisms that lead to superconducting states 
differing from the conventional Bardeen-Cooper-Schrieffer (BCS) paradigm~\cite{C56,BCS57a,BCS57b}, as well as a complete understanding of the pseudogap state \cite{TS99,NPK05} has driven the emergence of elaborate theoretical concepts that well transcend the goal of understanding any particular material. \\
There is strong evidence that suggests that interactions are mainly responsible for, or play a major role in generating the rich phase diagrams of these systems \cite{GLM12,vishik:2012, NAF14,CFY14,hashimoto:2014, kaminski:2014}. Under the umbrella of such an observation, many novel interesting ideas have emerged that were later found to be relevant also in other, completely unrelated systems. The marginal Fermi liquid~\cite{VLS89,V97,VNS02} introduced phenomenologically to account for experimental observations~\cite{TOR88} is a prime example. It was argued that strong interactions could cause the imaginary part of the self-energy, usually associated with the quasiparticle life-time, to behave as $\mathrm{Im}\Sigma(\omega)\sim\omega$ instead of the quadratic Fermi-liquid like behaviour \cite{AGD07}. Remarkably, this particular idea was also shown to be relevant for the physics of graphene~\cite{GGV99,GVV09}.   \\
On the other hand, high-energy physics has also profited from similar phenomenological approaches predicting a
plethora of verifiable consequences for experiments at the Large Hadron Collider (LHC)~\cite{Z10}. A particularly appealing idea, proposed recently by Georgi \cite{G07,G07b}, is the existence of a conformally invariant sector that couples to ordinary standard model particles. The former sector does not behave as ordinary matter since it is described by propagators without poles, and hence has no \emph{a priori} particle interpretation. Such unparticle "stuff" (as dubbed by Georgi) could in principle leave a very particular signature in LHC scattering cross-sections as missing spectral weight corresponding to a non-integer number of ordinary particles \cite{G07,CKY07}.
The parallelism with non-Fermi liquid properties~\cite{VNS02} recently motivated the idea that unparticles resulting from strong interactions could be responsible for the missing spectral weight in the pseudogap phase~\cite{phillips:2013}, accounting also for a possible breakdown of Luttinger's theorem~\cite{dave:2013}. 
These works complement the numerous scenarios that can lead to exotic superconductivity proposed in the literature, which have explored the dependence of the superconducting transition temperature with model parameters~\cite{KS90,KHV11,YHH14,MMS14}.
In the case of unparticles, within a BCS framework it was shown~\cite{phillips:2013} that if electrons participating in superconductivity were promoted to unparticles that pair with a standard BCS interaction, unusual phenomenology followed. 
 \\
Inspired by these ideas, we aim to discuss a mechanism which is distinct from other exotic forms of superconductivity and which can in principle exist in strongly correlated materials.  In this work we explore how superconductivity of normal electrons can arise from \emph{mediating} unparticles. In essence, we discuss how the unparticle construction provides a path from a high-energy theory with repulsive interactions between fermions to an effective low energy theory where particles attract. The price to pay is that the mediating glue is composed of unparticles, ultimately resulting in unconventional (i.e.~away from standard BCS) behaviour. Being a generic theoretical scenario, this mechanism could potentially emerge in strongly correlated matter. 
In this context, this work aims to clarify two important issues (i) the simplest theoretical construction where this effective unparticle mediating interaction emerges (ii) how this state departs from typical BCS phenomenology.\\
To address these points, we will first discuss a minimal theoretical framework
for the mechanism to emerge, partially reviewing known properties of unparticles, but focusing on those relevant for our work. 
We will write down a generic model containing the unparticle analogue of the Coulomb interaction
for which we write and solve the gap equation at the mean-field level and $s$-wave pairing. 
We will find that even at the mean field level, the unparticle nature of the mediators manifests itself through non-BCS phenomenology. We finish with a discussion of our results and a summary of our main conclusions. 

\begin{figure}
\centering
\includegraphics[scale=0.23]{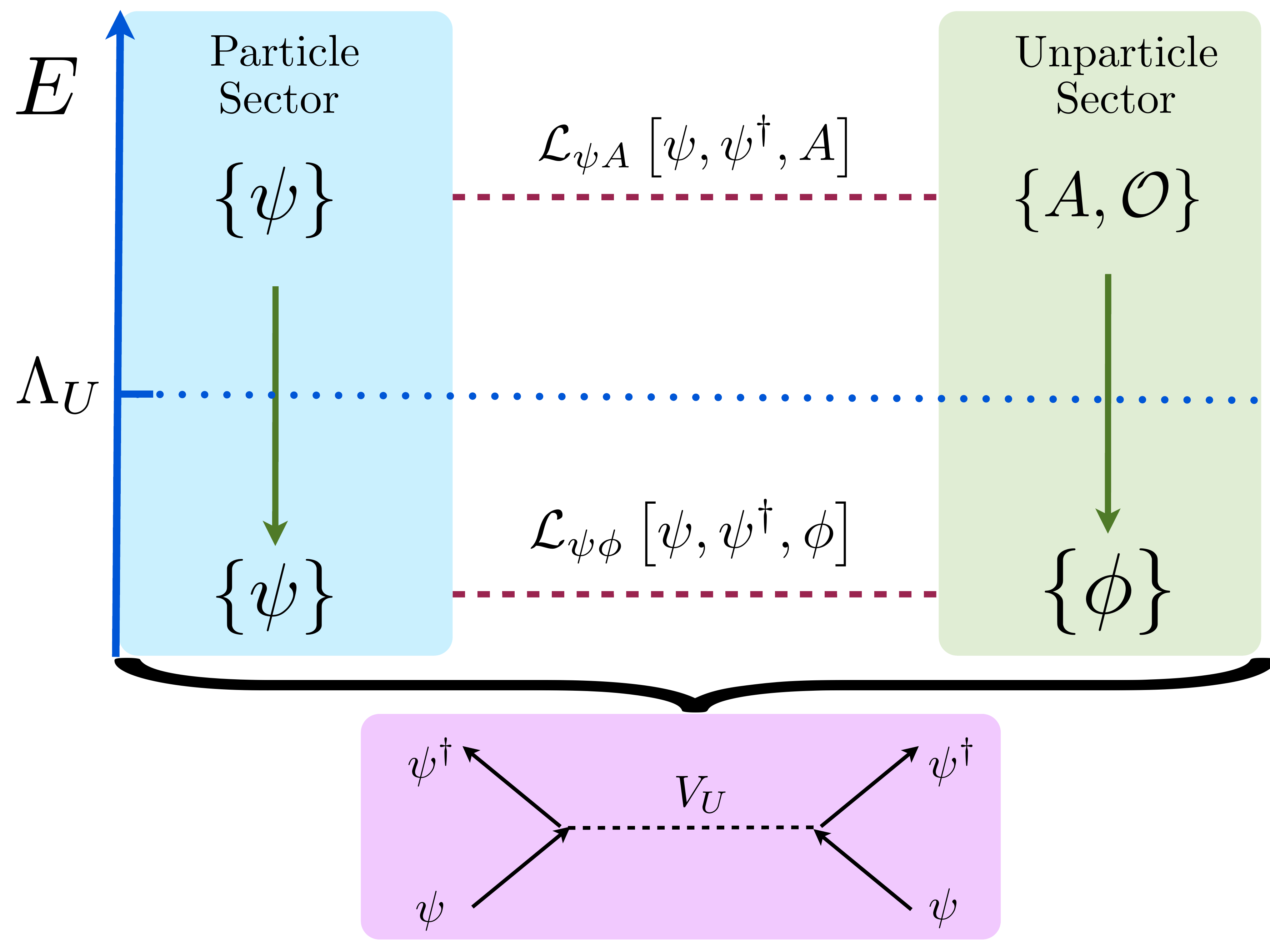}
\label{fig:schematic}
\caption{\label{fig: schematic} (Color online) Schematic picture showing the emergence of unparticles. At high energies ordinary particles (particle sector) couple to a second sector, the unparticle sector, that has a non-trivial infrared fixed point below a characteristic scale $\Lambda_{U}$ where the fields turn effectively into unparticles. Below $\Lambda_{U}$ this sector couples to the particle sector that is unaffected by the RG flow by construction. This coupling results in an effective four point interacting potential $V_{U}$, the unparticle analogue of the Coulomb interaction, defined by \eqref{eq:potential} [see main text].}
\end{figure}

\section{Emergence of unparticles} Although the concept of unparticles may seem exotic at first, especially 
in the context of strongly correlated electrons, it is not entirely novel to condensed matter systems. 
Within the renormalization group approach and due to electron interactions, quasiparticle propagators 
can acquire anomalous dimensions near non-trivial fixed points~\cite{LB91, G07}, 
turning them into unparticles. As discussed in \cite{phillips:2013} the Tomonoga-Luttinger liquid exhibits such behaviour~\cite{T50,L63,S11}. 
A second particularly clean, and to some extent unexplored example in $2+1$ dimensions that illustrates how unparticles
emerge is that of the low energy interacting electrons in graphene. 
In this system interactions renormalize the Fermi velocity $v_{F}$~ \cite{GGV94,KUP12} which increases at low energies, a prediction consistent with recent experiments \cite{EGM11}. 
However, as $v_{F}$ increases, the ratio with the speed of light $v_{F}/c$ ceases to be small and thus one has to consider the full relativistic Coulomb interaction to have a fully consistent theory \cite{GGV94}. In this case, the theory can be shown to have a non trivial infra-red (IR) fixed point at $v_{F}=c$. Close to this point, the electron propagator acquires an anomalous dimension $\gamma$ and satisfies the scaling law
\begin{equation}\label{eq:unparprop}
G(\lambda \omega,\lambda k) = \lambda^{\gamma} G(\omega, k),
\end{equation}
where $\lambda$ is the scaling parameter and $\gamma$ is the anomalous dimension, which is in general not an integer and depends on the coupling constant at the non-trivial critical point. For graphene $\gamma=\alpha^2/(12\pi^2)-1$ where $\alpha=e^2/\hbar c$ is the fine structure constant~\cite{GGV94}. Such a propagator corresponds by definition to an unparticle since it is defined around a fixed point (i.e.~it is conformally invariant) and has no simple quasiparticle poles in general. Thus electrons in graphene can be ultimately reinterpreted as genuine unparticles below an energy scale $\Lambda_{U}$ \footnote{The energy scale at which this happens is extremely small, albeit not zero, see \cite{JGV10}.}. \\
Motivated by this observation, we now reinterpret Georgi's original construction,\cite{G07} where an unparticle sector is coupled to ordinary particles, to obtain a theory that will lead us to unparticle mediated superconductivity. The process, summarized in Fig.~\ref{fig: schematic} starts by considering a theory with Lagrangian
\begin{subequations}
\begin{eqnarray}\label{eq:lagUV}
\mathcal{L}\left[\psi^{\dagger},\psi,A,\mathcal{O}\right]  &=& \mathcal{L}_{1}\left[\psi^{\dagger},\psi,A\right]+ \mathcal{L}_{2}\left[A,\mathcal{O}\right],\\
\label{eq:lagUV1}
\mathcal{L}_{1}\left[\psi^{\dagger},\psi,A\right]  &=&\mathcal{L}_{\psi}\left[\psi^{\dagger},\psi\right] +\mathcal{L}_{\psi A}\left[\psi^{\dagger},\psi,A\right],\\ 
\label{eq:lagUV2}
\mathcal{L}_{2}\left[A,\mathcal{O}\right]  &=& \mathcal{L}_{A}\left[A\right]  +\mathcal{L}_{A\mathcal{O}}\left[A, \mathcal{O}\right]  +\mathcal{L}_{\mathcal{O}}\left[\mathcal{O}\right]  ,
\end{eqnarray}
\end{subequations}
defined for energies $E \gg \Lambda_{U}$. The theory is composed of two sectors which we illustrate schematically in Fig.~\ref{fig:schematic}. The first sector, $\mathcal{L}_{1}\left[\psi^{\dagger},\psi,A\right]$, represents ordinary particles $\psi$ (e.g.~electrons) coupled through a generic interaction $A$, (e.g.~the Coulomb interaction). 
In the original unparticle set up, this sector would correspond to the standard model of particle physics. 
On the other hand $\mathcal{L}_{2}\left[A,\mathcal{O}\right]$ represents a different interacting sector where other degrees of freedom, 
collectively denoted by the field $\mathcal{O}$, interact also through $A$. At this point we impose that $\mathcal{L}_{2}\left[A,\mathcal{O}\right]$ has a non trivial IR fixed point below the energy scale $\Lambda_{U}$. Above this energy scale $\mathcal{L}_{2}\left[A,\mathcal{O}\right]$ has an ordinary interpretation as a particle theory. Below $\Lambda_{U}$ unparticles emerge as exemplified at the beginning of the section (see Fig.~\ref{fig:schematic}). 
We emphasize that the scale $\Lambda_{U}$ must exist in order to constrain the unparticles to exist \emph{only} 
below this energy scale. Remarkably, there are explicit examples where it is possible to estimate this energy scale (see for instance eq. (3) in \cite{georgi:2008} or eq. (23) in \cite{JGV10}).
\\
In particular, if $\mathcal{L}_{1}$ is unaffected by the IR renormalization group (RG) flow, at energies lower than $\Lambda_{U}$, ordinary particles $\psi$ will effectively only couple to unparticles \cite{G07}, which we label $\phi$ with an action of the form
\begin{equation}\label{eq:lagIR}
\mathcal{L}_{IR} =  \mathcal{L}_{\psi}+\mathcal{L}_{\psi \phi} + \mathcal{L}_{\phi}.
\end{equation}
So far, we have just reviewed how unparticles generically emerge. 
We now assume that $\mathcal{L}_{IR}$ is quadratic in the fields and 
integrate out $\phi$ to obtain an effective interaction between electrons, mediated by an
unparticle propagator of the form~\eqref{eq:unparprop}. After this step 
the Lagrangean reads
\begin{equation}\label{eq:Ulag}
\mathcal{L}_{IR} =  \mathcal{L}_{\psi} + \dfrac{g}{2}\sum_{ij}\psi^{\dagger}_{i}\psi_{i} G_{\phi; ij}\psi^{\dagger}_{j}\psi_{j} ,
\end{equation}
where $g$ is a coupling constant and $G_{\phi;ij}$ is the unparticle propagator in real space, which is in general a matrix depending on the physical degrees of freedom encoded in $\psi$. 
Generically, the functional form of the propagator $G_{\phi;ij}$ is fixed
by dimensionality and conformal invariance of the unparticle sector at the IR fixed point. In particular for a scalar unparticle 
the propagator is of the form~\cite{G07,GS08} 
\begin{eqnarray}
G_{\phi} (q)&=&\dfrac{A_{d_{U}}(q^2)^{d_{U}-2}}{2(\Lambda_{U}^2)^{d_{U}-1}\sin(\pi d_{U})},\\
A_{d_{U}}&=&\dfrac{16\pi^{5/2}}{(2\pi)^{2d_{U}}}\dfrac{\Gamma(d_{U}+\frac{1}{2})}{\Gamma(d_{U}-1)\Gamma(2d_{U})}, 
\end{eqnarray}
Comparing to Eq.  \eqref{eq:unparprop},  $\gamma = 2(d_{U}-2)$. The quantity $d_{U}$ is referred to as the unparticle scaling dimension. 
For a scalar particle it satisfies $d_{U}\geq 1$ to preserve unitarity of the theory \cite{Mack77}. We are interested in the unparticle generalization of the Coulomb interaction and thus we focus on  the effective static potential $V_U$ associated to this scalar propagator which was derived in Refs.~\cite{GS08,GN08}. This is
a good approximation when retardation effects associated to field $\phi$ (i.e. current-current interactions) can be neglected, which implies
that the typical Fermi velocity should satisfy $v_F/c\ll1$ \cite{HNP73}. In this approximation 
the effective static potential reads
\begin{eqnarray}\nonumber
V_{U} (\mathbf{q})&=&4\pi g \dfrac{A_{d_{U}}}{2(\Lambda_{U}^2)^{d_{U}-1}\sin(\pi d_{U})}(\mathbf{q}^2)^{d_{U}-2}\\
\label{eq:potential}
&\equiv &C_{d_{U}}(\mathbf{q}^2)^{d_{U}-2}.
\end{eqnarray}
Note that mathematically, and for $1\leq d_{U}<2$, it is possible to rewrite this potential to be proportional to the integral $\int^{\infty}_{0}dM^2\rho(M^2,d_{U})/(\mathbf{q}^2+M^2)$ with $\rho(M^2,d_{U})=(M^2)^{d_{U}-2}$.
This form explicitly reveals that unparticles can be interpreted as a 
tower of infinitely many massive particles distributed according to $\rho(M^2,d_{U})$~\cite{K07,S07,DH08}.
Indeed, when examined in real space, this potential is nothing but a Yukawa potential integrated over the screening momentum scale $M^2$. Physically, it is possible to interpret such an interaction as critical in the sense that it has the form of a screened interaction, but the screening occurs at all length scales, effectively resulting in a potential of the form $V_{U} (\mathbf{r})\propto 1/\left|\mathbf{x}\right|^{2d_{U}-1}$~\cite{GN08,GS08}, the unparticle counterpart of the Coulomb interaction \footnote{Technically, the integral over $M^2$ only converges for $1<d_{U}<2$. For the same interpretation to hold for $d_{U}>2$ a ultraviolet cut-off must be provided for the the integral, which is on the other hand natural for condensed matter systems}. Interactions of this form have also been discussed in the context of the even denominator fractional quantum Hall states~\cite{HLN93,NW94,MMS14}. From Eq.~\eqref{eq:potential} the long range Coulomb interaction is recovered for $d_{U}=1$ if we identify $g=e^2$.  \\
A crucial observation here is that the sign of $g$, and thus the attractive or repulsive character of the interaction is not fixed by the renormalization that generates the unparticles, described above~\cite{KL65}. A repulsive interaction between fermions $\psi$ at high energies in \eqref{eq:lagUV1}, represented by the term $\mathcal{L}_{\psi A}\left[\psi^{\dagger},\psi,A\right]$ can turn into an effective attractive interaction mediated by unparticles at low energies. The actual sign will depend on the nature of the IR fixed point itself and the particular value of $d_{U}$ trough the prefactor $C_{d_{U}}$. We use this freedom to fix the low energy theory to have an attractive character.\\
We finish this section with a general remark on the validity of the present approach. 
In a general scenario there can exist an additional energy scale, $\Lambda_e$, below which the field 
$\psi$, acquires anomalous dimensions, and is described by an unparticle propagator. 
In this work we assume that there exists an energy window such that $ \Lambda_{e}<E<\Lambda_{U}$
where the field $\psi$ is still defined by quasiparticle propagators that interact via the unparticle potential 
\eqref{eq:potential}. 
In this notation, Ref.~\cite{phillips:2013} can be interpreted as the case where $ \Lambda_{U}<E<\Lambda_{e}$ 
and the scenario where $E< (\Lambda_{U},\Lambda_{e})$ is yet to be explored.\\

\section{Mean field approximation} 
\begin{figure}
  \begin{center}
      \includegraphics[width=0.7\linewidth]{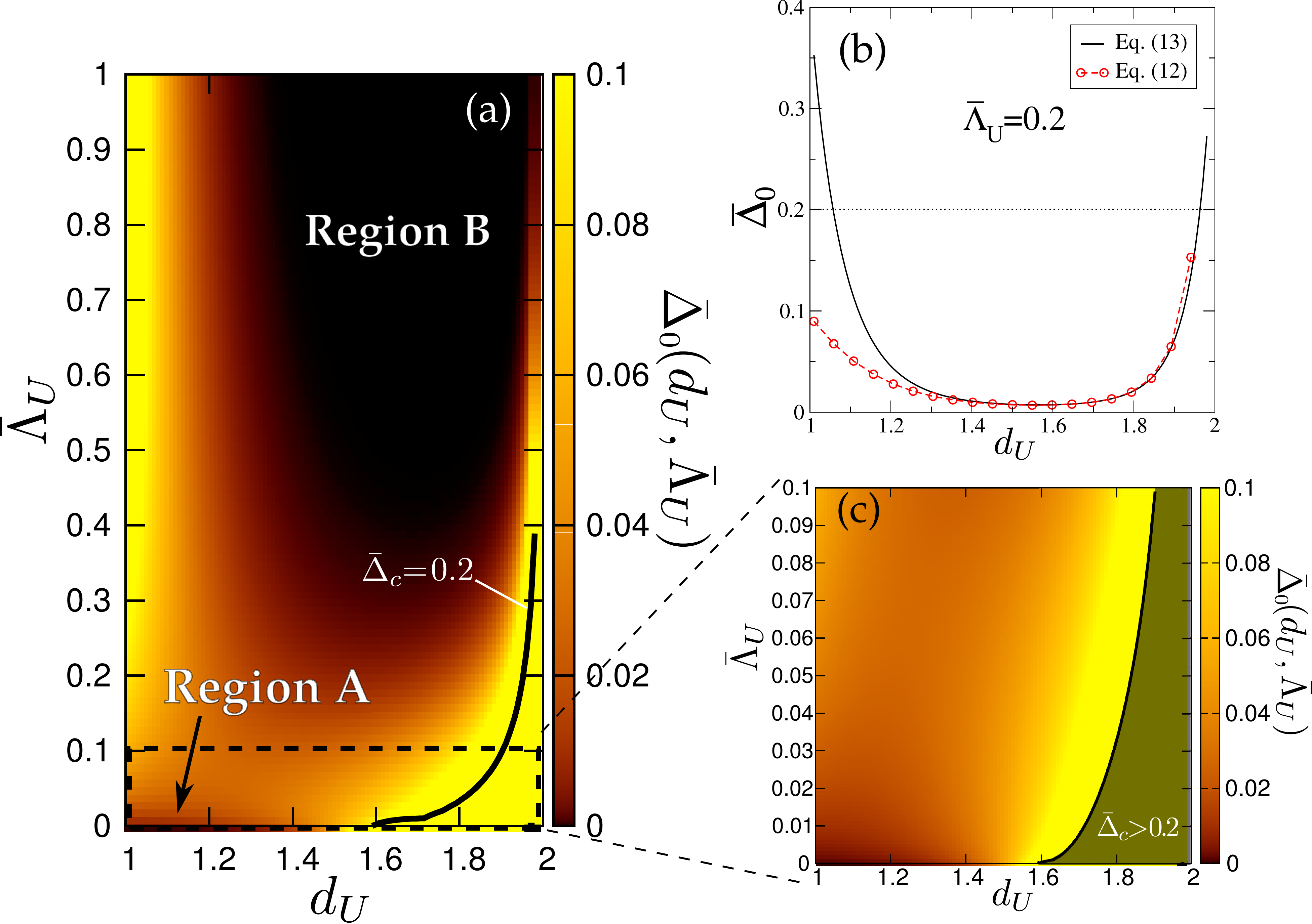}
  \end{center}
  \caption{\label{fig:gap}(Color online)(a) The gap in units of the Fermi energy $\bar{\Delta}_{0}$ as a function of $\bar{\Lambda}_{U}=\Lambda_{U}/k_{F}$ and $d_{U}$ calculated from Eq.~\eqref{eq:adimgap} with $gk_{F}/\varepsilon_{F}=1$ (see main text).  Solutions below the solid black line can be disregarded as unphysical since they satisfy $\bar{\Delta}_0>\bar{\Delta}_c=0.2$, where $\bar{\Delta}_{c} \ll 1 $ is chosen for illustrative purposes.  (b) Comparison of the analytic result of Eq.~\eqref{eq:onshell-limit} and the numerical result for $\bar{\Lambda}_U=0.2$ as a function of the unparticle dimension $d_U$. (c) Numerical plot of small $\bar{\Lambda}_U$ region of numerical computation shown in (a). The shaded region represents unphysical $\bar{\Delta}_0>\bar{\Delta}_c$ solutions. }
\end{figure}

Given the above discussion we now ask if the attractive unparticle Coulomb potential \eqref{eq:potential} can lead 
to unparticle mediated superconductivity, and if so, what is its particular signature.
In order to accomplish this, we consider the  case of coherent quasiparticles in the presence of scale invariant bosonic unparticles.  We then proceed in the spirit of a modified BCS-theory, which now contains a potential due to mediating unparticles which has a natural cutoff, $g\Lambda_U$, the highest energy an unparticle might have.
We proceed without explicit interactions, but keep in mind that we are free to consider any renormalized quasiparticle weight by including $Z_k$ factors in the effective mass.
Switching to the equivalent Hamiltonian formalism the effective low energy Hamiltonian follows from
\eqref{eq:Ulag}
\begin{equation}\label{eq:Ham}
H= \sum_{\vect{k}} \xi_{\vect{k}}\psi^{\dagger}_{\vect{k}}\psi_{\vect{k}}
+\sum_{\vect{kk'q}} V_{U}(\mathbf{q})\psi^{\dagger}_{\vect{k}}\psi_{\vect{k-q}}\psi^{\dagger}_{\vect{k'+q}}\psi_{\vect{k'}},
\end{equation}
where the sums are a short hand notation for three-dimensional momentum space integrals. The first term corresponds to the diagonalized Hamiltonian stemming from $\mathcal{L}_{\psi}$ in \eqref{eq:Ulag}  with $\xi_{\vect{k}}=\varepsilon_{\vect{k}}-\mu$ with $\mu$ being the chemical potential and $\varepsilon_{\vect{k}}$ a dispersion relation (fixed below). The second part contains the unparticle mediated interaction, which we set to be attractive by choosing the sign of $C_{d_{U}}$ or the appropriate $d_{U}$ regime. \\
For concreteness we choose $\psi$ to describe fermions in three spatial dimensions with quadratic dispersion $\varepsilon_{\vect{k}}=\vect{k}^2/2m$. Following standard techniques \cite{BF07} we can write down the self consistent mean field equation for the superconducting gap $\Delta_\vect{k}$ as
\begin{equation}\label{eq:gapeq}
\Delta_\vect{k} = - \sum\limits_{\vect{k'}} \frac{V_{U}(\vect{k}-\vect{k'}) \Delta_{\vect{k'}}}{ 2 \sqrt{\xi_{\vect{k'}}^2+\Delta_{\vect{k'}}^2}} \left[ 1- 2n_F(E_{\vect{k'}}) \right],
\end{equation}
where $n_F(E_{\vect{k}})$ is the Fermi-Dirac distribution function and $E_{\vect{k}}=\sqrt{\xi^2_{\vect{k}}+\Delta^2_{\vect{k}}}$. 
Unparticle physics enters through the potential $V_{U}(\vect{k}-\vect{k'})$. We look for an isotropic
$s$-wave solution for the superconducting gap and thus we impose that $\Delta_\vect{k}=\Delta_{0}$.
It is now possible to integrate the angular variables in \eqref{eq:gapeq}, resulting in a gap equation of the form \eqref{eq:gapeq}
but with the effective potential (see~\ref{sec:APP})
\begin{equation}
\label{eqn:vfirst}
V_{\mathrm{eff}}(k,k') = -\dfrac{|C_{d_U}|}{4\pi(d_{U}-1){kk'}}\sum_{s=\pm}s(k+sk')^{2d_{U}-2},
\end{equation}
that depends only on the absolute values of both $k=|\vect{k}|$ and $k'=|\vect{k'}|$. In order
to solve the gap equation with this potential, we make use of the standard 
Fermi surface restricted (FSR) approximation~\cite{MCC14} by setting $k=k_{F}$ to obtain,
at $T=0$
\begin{eqnarray}
1&=&  -\int \dfrac{k^2dk}{2\pi}  \frac{V_{eff}(k_{F},k) }{2 \sqrt{\xi_{k}^2 + \Delta_{0}^2}}.
\end{eqnarray}
At this point we find it convenient to integrate over the energy variable 
$\varepsilon_{\vect{k}}$ and write the equation in a more transparent dimensionless form by defining $\bar{k}=k/k_{F}$,
$\bar{\Delta}_{0}=\Delta_{0}/\varepsilon_{F}$, $\bar{\epsilon}=\bar{k}^2-1$.
After a few lines of algebra we obtain 
\begin{eqnarray}
1= \dfrac{2m (k^2_{F})^{d_{U}-3/2}|C_{d_{U}}|}{ 16 \pi^2 (d_{U}-1)} 
\label{eq:adimgap}
\int d\bar{\epsilon}\dfrac{\sum\limits_{s=\pm} s\left[\left(1+s\sqrt{\bar{\epsilon}+1}\right)^2\right]^{d_{U}-1}}{2 \sqrt{\bar{\epsilon}^2 + \bar{\Delta}_{0}^2}}.
\end{eqnarray}
The integral is evaluated in an energy shell around the Fermi surface defined by the inequality $|\bar{\epsilon}|< g\Lambda_{U}/\varepsilon_{F}$, where $g\Lambda_{U}$ is the typical energy scale under which the mediating interaction due to unparticles is non-zero\footnote{We use units where $g$ has dimensions of energy$\times$(momentum)$^{-1}$ and $\Lambda_{U}$ has units of momentum. Thus, $g\Lambda_{U}$ is the natural energy scale for unparticles.}. \\
Note that the density of states (DOS) at the Fermi level of a 3D electron gas is $\rho(\varepsilon_{F})^2\propto k_{F}^2$ and therefore the multiplicative prefactor in front of the integral is proportional to $\left[\rho(\varepsilon_{F})^2\right]^{d_{U}-3/2}$. This is in sharp contrast with conventional BCS where the equivalent prefactor is proportional to $\rho(\varepsilon_{F})$~\cite{BF07}. The latter is recovered for $d_{U}\to 2$ since the interaction \eqref{eq:potential} approaches the short range BCS like interaction. We emphasize in particular that for $d_{U}=3/2$ the dependence on the DOS at the Fermi level drops out \emph{completely} from the gap equation. This case is special in that it corresponds to a $1/\mathbf{x}^2$ interaction which in 3+1 dimensions scales exactly as the kinetic term~\cite{Landauvol3}. \\
We continue the comparison with the BCS scenario by employing an approximation where we only allow particles to exchange momentum 
in a infinitesimal shell close to the Fermi surface. 
In this approximation we are able to set $\bar{k}=1$ $(\bar{\epsilon}=0)$ in the numerator of
\eqref{eq:adimgap}. By solving the remaining integral we arrive at the gap formula 
\begin{eqnarray}\label{eq:onshell-limit}
\Delta_{0}=2g\Lambda_{U}\mathrm{exp}\left[-1/\tilde{V}\rho(\varepsilon_{F})^{2d_{U}-3}\right],
\end{eqnarray}
with an effective interaction $\tilde{V}=\frac{|C_{d_{U}}|}{d_{U}-1}\left[\frac{4\pi^2}{2m}\right]^{2d_{U}-4}$. In this format it is tempting to interpret $g\Lambda_{U}$ as the analogue of the Debye frequency in BCS superconductivity \cite{BF07} since Eq.~\eqref{eq:onshell-limit} approaches the standard BCS result as $d_{U}\to 2$. However, the BCS character is lost in general because the exponential factor in Eq. \eqref{eq:onshell-limit} depends both on $d_{U}$ and $\Lambda_{U}$, the latter entering through $\tilde{V}$. Indeed, $C_{d_{U}}\propto 1/\Lambda_{U}^{2d_{U}-2}$ that results in a gap dependence $\Delta_0 \propto \Lambda_U \mathrm{exp}\left[ - \Lambda_U^{2d_U-2} \right]$ that has no BCS analogue. \\
The fact that both the unparticle energy scale and the effective interaction strength are functions of $\Lambda_U$ allows for distinctly non-BCS solutions of~\eqref{eq:adimgap}. To prove that these solutions exist we have solved~\eqref{eq:adimgap} numerically for  $\bar{\Delta}_0$ with no implicit assumption on $k^\prime$ or the size of $\Lambda_U$. In order to proceed, note that the prefactors in Eq.~\eqref{eq:adimgap} can be rewritten as $\frac{A_{d_U}}{8\pi (d_U-1)\sin(\pi d_U)}\frac{g k_F}{\varepsilon_F}\frac{1}{\left(\bar{\Lambda}_{U}^2\right)^{d_U-1}}$ where $\bar{\Lambda}_U=\Lambda_U/k_F$.  If we  assume that $g$ is of the order of the 
bare Coulomb interaction then the dimensionless ratio $g k_F/\varepsilon_F$ is of order one for typical values of $k_{F}$ and $\varepsilon_{F}$. Thus we are able to focus on the effect of increasing the shell around the Fermi surface defined by $\bar{\Lambda}_{U}$ as well as the effect of $d_{U}$. Furthermore, since the superconducting gap should be smaller than $\varepsilon_{F}$, it should satisfy $\bar{\Delta}_0\ll 1$. \\
Our results are summarized in Fig.~\ref{fig:gap}(a) where we plot the numerical solution of $\bar{\Delta}_0$ as a function of  $\bar{\Lambda}_{U}$ and $d_{U}$.  It is possible to distinguish two regions where $\bar{\Delta}_0$ vanishes labelled $A$ and $B$.  
We find that regions $A$ and $B$ are separated by robust physically acceptable solutions where  $\bar{\Delta}_0\ll 1$ over a wide range of $d_U$ and $\bar{\Lambda}_U$. In Fig.~\ref{fig:gap}(b) we show an example of such an acceptable region for $\bar{\Lambda}_{U}=0.2$. The two curves compare solutions obtained with \eqref{eq:adimgap} and \eqref{eq:onshell-limit} showing that the latter is an increasingly better approximation as $d_{U}\to 2$. \\
In contrast, we find that, as $\bar{\Lambda}_{U} \to 0$ and $d_{U}>3/2$ [see Fig.~\ref{fig:gap}(c)]. the approximations that lead to \eqref{eq:onshell-limit} break down (for instance $\tilde{V}$ ceases to be small) and \eqref{eq:adimgap} leads to solutions beyond the upper bound $\bar{\Delta}_{0}=1$, which diverge when $\bar{\Lambda}_U \to 0$ and $d_{U}\to 2$. As $d_{U}\to 1$ the gap decreases leading once more to physically valid solutions.\\
All the discussed solutions are only possible because both the effective interaction strength and the upper limit for the unparticle energies are set by $\Lambda_{U}$ and the two effects can level out at a given value of $d_{U}$ to generate a physically acceptable solution for the superconducting gap. Together with their dependence on physical parameters discussed above, their emergence is in sharp contrast to how standard BCS solutions occur. Thus, their existence and scaling with $\Lambda_{U}$ and $d_{U}$ represents one of the main results of this work.
\section{Discussion and conclusions}

In this work we have described how it is possible for unparticles in strongly
correlated matter to mediate superconductivity. We have shown that this scenario
can depart strongly from the conventional BCS paradigm. In particular, we have found that for $s-$wave superconductivity, 
the superconducting gap acquires an unusual dependence on the DOS at the Fermi level that can even vanish
when the scaling dimension of the mediating unparticle is $d_{U}=3/2$.  
Interestingly, this vanishing dependence on the density of states occurs at the value of $d_U$ where kinetic and potential energy terms scale equally in energy\cite{Landauvol3}.  This may be important in the high-$T_c$ materials which are thought to switch from typical potential energy driven pairing in the overdoped side of the phase diagram, to kinetic energy driven in the underdoped\cite{basov:1999, gull:2012}.  Such a crossover may correspond to a similar crossover of the unparticle scaling determined by the value of $d_U$.
\\
As further exemplified above with electrons in graphene, obtaining $d_{U}$ for a particular condensed matter system is sometimes possible.  However, this is a hard task in general since it implies characterizing the crossover from the high-energy particle description to unparticles, which could be dominated by a strong coupling IR fixed point hard to access analytically~\cite{G07,G07b}. The particular value of $d_{U}$ will generically depend on the coupling constants in the unparticle sector (that differ from $g$ in general)~\cite{GGV94,G07} and can therefore scale with an external experimentally tunable parameter that modifies the interaction strength. This idea resembles the situation in hole-doped high-$T_c$ materials that can be doped from strongly correlated Mott insulators towards Fermi liquids for large hole doping~\cite{norman:2003}. The presented dependence on $d_U$ can significantly affect the size of the gap as shown in Fig.~\ref{fig:gap}(b).
 \\
To conclude, our work also leaves many interesting open extensions such as unparticle mediated $d$-wave superconductivity and the effects of reduced dimensionality or competing orders. Furthermore, the presented construction may be useful to generalize previous studies in the context of even denominator fractional quantum Hall state theories~\cite{HLN93,NW94,MMS14}. The exact connection to the latter and its consequences will be explored in future work.\\

\appendix
\section{Effective Potential}\label{sec:APP}

In this section we present the derivation of the gap formula and effective potential used in 
the main text. We start from the standard gap equation (9) given in the main text
\begin{equation}
\Delta_\vect{k} = - \sum\limits_{\vect{k'}} \frac{V_{U}(\vect{k}-\vect{k'}) \Delta_{\vect{k'}}}{ 2 \sqrt{\xi_{\vect{k'}}^2+\Delta_{\vect{k'}}^2}} \left[ 1- 2n_F(E_{\vect{k'}}) \right],
\end{equation}
we are most interested in isotropic $s-$wave pairing and thus set $\Delta_{\vect{k}}=\Delta_{0}$.
This enables us to perform the angular integral first that defines the effective potential
\begin{equation}
V_{\mathrm{eff}}(\vect{k},\vect{k'}) = \int\dfrac{d\theta_{\vect{k'}} d\phi_{\vect{k'}}}{(2\pi)^2}\sin(\theta_{\vect{k'}})  V_{U}(\vect{k}-\vect{k'}).
\end{equation}
In the isotropic $s-$wave pairing case we can fix the direction of $\vect{k}$ to conveniently solve the integral
\begin{equation}
V_{\mathrm{eff}}(k,k') = \dfrac{C_{d_U}}{2\pi} \int\limits_0^{\pi} \left( k^2 +{k'}^2 -2k {k'} \cos(\theta') \right)^{du-2} \sin(\theta') d\theta',
\end{equation}\normalsize
which results in
\begin{eqnarray}
V_{\mathrm{eff}}(k,k') &=& \dfrac{C_{d_U}}{2\pi} \Bigg[ \frac{(k+{k'})^{2d_{U}-2}-(k-{k'})^{2d_{U}-2}}{2k{k'}(d_U-1)}   \Bigg],\\
&=&\dfrac{C_{d_U}}{4\pi(d_{U}-1){kk'}}\sum_{s=\pm}s(k+sk')^{2d_{U}-2},
\end{eqnarray}
valid for $1<d_{U}<2$ presented in the main text. To solve the remaining integral we make use of the 
Fermi surface restricted (FSR) approximation (see for instance~\cite{MCC14}) and set $k=k_{F}$.
For this case, and $T=0$, the gap equation reads
\begin{eqnarray}
1&=&  \int \dfrac{k^2dk}{2\pi}  \frac{V_{eff}(k_{F},k) }{2 \sqrt{\xi_{k}^2 + \Delta_{0}^2}},
\end{eqnarray}
where we have dropped the primes. The effective potential, used in the main text is given therefore by
\begin{eqnarray}
V_{\mathrm{eff}}(k_{F},\bar{k})=\dfrac{C_{d_U}(k^2_{F})^{d_{U}-2}}{4\pi(d_{U}-1){\bar{k}}}\sum_{s=\pm}s(1+s\bar{k})^{2d_{U}-2},
\end{eqnarray}
with $\bar{k}=k/k_{F}$.

\vspace{12pt}

\ack
A.G.G. acknowledges F. de Juan and M. A. H. Vozmediano 
for related collaborations from which this work originated as well as critical reading of the manuscript. 
We are indebted to Jens H. Bardarson, E. V. Castro, P. Fulde and B. Valenzuela for useful comments and 
M. Peiro for discussions at the early stages of this work. 
\vspace{12pt}

%\bibliographystyle{new2}
%\bibliography{UPSC_2.bib}

\end{document}